\documentclass[referee]{raa}
\pdfoutput=1
\usepackage{graphicx,times}
\usepackage{natbib}
\usepackage{amssymb,amsmath}
\usepackage{ulem}
\usepackage[thicklines]{cancel}
\usepackage{xcolor}

\bibpunct{(}{)}{;}{a}{}{,}

\usepackage[pagebackref=true]{hyperref}

\begin{document}

  \title{Testing the coincidence problem with strong gravitational lens, Type Ia supernovae and Hubble parameter observational data
}

   \volnopage{Vol.0 (20xx) No.0, 000--000}
   \setcounter{page}{1}

   \author{JingWang Diao
      \inst{}
   \and Yu Pan\footnote{Corresponding author}
      \inst{}
   \and Wenxiao Xu
      \inst{}
   }

   \institute{School of Science, Chongqing University of Posts and Telecommunications, Chongqing 400065, China; {\it panyu@cqupt.edu.en}\\
\vs\no
   {\small Received 20xx month day; accepted 20xx month day}}

\abstract{ In this paper, we use three different kinds of observational data, including 130 strong gravitational lensing (SGL) systems, type Ia supernovae (SNeIa: Pantheon and Union2.1) and 31 Hubble parameter data points ($H(z)$) from cosmic chronometers to constrain the phenomenological model ($\rho_x\varpropto\rho_m a^{\xi}$). By combining these three kinds of data (Union2.1+SGL+$H(z)$), we get the parameter value at the confidence interval of $2\sigma$, $\Omega_{X,0} = 0.69\pm0.34$, $\omega_x = -1.24\pm0.61$, $\xi = 3.8\pm3.9$ and $H_0 = 70.22\pm0.86$ kms$^{-1}$Mpc$^{-1}$. According to our results, we find that the $\Lambda$CDM model is still the model which is in best agreement with the observational data at present, and the coincidence problem is not alleviated. In addition, the $\Omega_X$ and $\Omega_m$ have the same order of magnitude in $0<z<1.26$. At last, we obtain the transition redshift $z_T=0.645$. If the transition occurs in $z>0.645$, it is necessary to introduce the dark energy interacting with dark matter.
}

   \authorrunning{JingWang Diao, Yu Pan and Wenxiao Xu }            
   \titlerunning{Testing the coincidence problem}  

   \maketitle

%
%
\section{Introduction}           
\label{sect:intro}

In recent years, cosmological observations such as baryon acoustic oscillation (BAO) (\citealt{Eisenstein2005}), type Ia supernovae (SNeIa) (\citealt{Riess+etal+1998}) and cosmic microwave background (CMB) (\citealt{Spergel+etal+2003}) suggest that the universe is in an accelerated state of expansion. The accelerating expansion of the universe has attracted widespread attention, and dark energy has been introduced as a universe component to explain this phenomenon. Currently, the composition of the universe given by the CMB is $68.89\%$ (\citealt{Aghanim+etal+2020}) for dark energy (DE) and $26.0\%$ (\citealt{Bennett+etal+2013,Ade+etal+2016}) for dark matter (DM). We can see that dark energy is currently in the driver's seat, determining the future of the universe. However, we now know nothing about the nature of dark energy. Therefore, people have done a lot of work on dark energy (\citealt{Spergel+etal+2003,Astier+etal+2006,Amanullah+etal+2010,Cao+etal+2011,Hicken+etal+2009,Gong+etal+2012}), and some models of DE have been proposed, including the $\Lambda$ cold dark matter model ($\Lambda$CDM) (\citealt{Carroll+etal+1992,Riess+etal+1998,Peebles+etal+2003}).

Cosmological observations suggest that the $\Lambda$CDM model almost agrees with all observational results, making it the best model currently available for describing the universe. However, there is a theoretical problem with this model, known as the cosmological constant problem (\citealt{Weinberg+etal+1989}). The cosmological constant problem includes two aspects: one is the coincidence problem (\citealt{Zlatev+etal+1999}) and the other is the fine-tuning problem. They are the vacuum energy density and matter density given by the cosmic observation are in the same order of magnitude; the theoretical value of the vacuum energy density is 120 orders of magnitude different from the observed value.

Since there is no convincing explanation for why DE predominates in the present, many possible theories can be used to mitigate the coincidence problem. What is interesting is that the interaction between DM and DE can be used to study the coincidence problem (\citealt{Amendola+etal+2000,CalderaCabral+etal+2009}). From the perspective of fundamental physics, people can not determine the specific form of interaction, so we can consider a variety of interaction models. We can choose the phenomenological combination of DE density and DM density to satisfy the premise by set them magnitude as the same order. DE and DM through the $Q$ to exchange energy, where $Q$ is the interaction term. Nowadays, there are many interacting dark energy (IDE) models, such as the $\xi IDE$ model, $\gamma_m IDE$ model and $\gamma_d IDE$ model. At the same time, many people have done research on the interaction of DE and DM. \citet{Chen+etal+2010} adopted BAO, SNe and CMB data to constrain the interaction model, and found that CMB data with high redshift may have more strict restrictions on $\xi$, where $\xi$ is the severity of the coincidence problem. By using the new GRB , Union2.1 SNe, CMB and BAO data sets, \citet{Pan+etal+2013} constrained an interaction dark energy model and obtained a slight conversion of dark matter to dark energy within a confidence interval. \citet{Pan+etal+2015} used $H(z)$, BAO and CMB data to constrain the interacting dark energy model, and the results show that $H(z)$ data can give a better limiting result for the interaction parameter $\gamma_m$. \citet{Lan+etal+2020} combined the high-redshift quasar (QSO) data with Union2.1 SNe data to constrain the $\xi$IDE model, and the results did not alleviate the coincidence problem.

As is known to all, in the $\Lambda$CDM model, the relation between matter density ($\rho_{m}$) and energy density ($\rho_{x}$) is $\rho_x\varpropto\rho_ma^3$ where $a=1/(1+z)$, and one would expect a $\rho_x\varpropto\rho_m$ relationship between DM and DE in a theory with no coincidence problem. In this work, we use a phenomenological model of relation $\rho_x\varpropto\rho_m a^\xi$ (\citealt{Dalal+etal+2001}) to research the coincidence problem, where $\xi$ indicates the severity of the coincidence problem. In this phenomenological model, $Q$ represents the exchange energy between DE and DM. If $Q=0$ ($\xi+3\omega_x =0$), where $\omega_x$ is the equation of state of DE, then the DE does not interact with DM in the standard cosmology. When $Q<0$ ($\xi+3\omega_x >0$), DM
transform into DE and the coincidence problem is not alleviated. When $Q>0$ ($\xi+3\omega_x <0$), DE transform into DM and the coincidence problem is alleviated. The SGL data satisfy the hypothesis of spherical symmetry in the lens mass model, and large amounts of SNe data have a stronger constraint effect. \cite{Amante+etal+2020} constrained the three models based on SGL data in conjunction with other lens data, and the results show that the cosmological parameters were very sensitive to the data they selected. \cite{Wang+etal+2020} used SGL and SNe to constrain the cosmic curvature, and the results show that the selection of lens model and the classification of SGL data can enhance the constraints on cosmic curvature. Since, in this work, we use type Ia supernova sample data (Pantheon and Union2.1) (\citealt{Scolnic+etal+2018,SupernovaCosmologyProject:2011ycw}), strong gravitational lensing (SGL) (\citealt{Chen+etal+2019}) and Hubble parameter ($H(z)$) (\citealt{Wei+etal+2017}) data to constrain the model parameters. In addition, the transition redshift ($z_T$) is also used to study the necessity of DE interacting with DM.

The layout of this article is as follows. In section \ref{sect:Obs}, we introduce the  phenomenological model equations. In section \ref{sect:data}, we submit cosmological observational data and analyze the results. In Section \ref{sect:analysis}, we introduce transition redshift and show the  results. Finally, in Section \ref{sect:discussion}, we made a summary and discussion.


\section{The phenomenological model equations}
\label{sect:Obs}
We use a phenomenological model of relation $\rho_x\varpropto\rho_m a^\xi$ (\citealt{Dalal+etal+2001}) to research the coincidence problem. In the flat FRW metric universe, the DE density parameter $\Omega_{X}$ and the DM density parameter $\Omega_{m}$ satisfy $\Omega_{X}+\Omega_{m}=1$, and $\Omega_{X}=\frac{\Omega_{X, 0} a^{\xi}}{1-\Omega_{X, 0}\left(1-a^{\xi}\right)}$, where $\Omega_{X,0}$ is the present value of $\Omega_{X}$. The conservation of energy equation can be written as
\begin{equation}
\frac{d \rho_{all}}{d a}=\frac{3}{a}\left(1+\omega_{x} \Omega_{X}\right) \rho_{all}=0
\end{equation}
where $\omega_{x}$ is the equation of state of DE, $\rho_{all}$ is the total density and $\rho_{all}=\rho_{x}+\rho_{m}$. In addition, the DE and DM do not evolve independently and exchange energy through interaction. We use $\rho_{x}=\kappa \rho_{m} a^{\xi}$ to obtain the following formula, where $\kappa$ is a constant
\begin{equation}
\frac{d \rho_{m}}{d a}+\frac{3}{a} \rho_{m}=Q,
\end{equation}
\begin{equation}
\frac{d \rho_{x}}{d a}+\frac{3}{a}\left(1+\omega_{x}\right) \rho_{x}=-Q,
\end{equation}
where $Q$ is the interaction term and $Q=-(\xi+3\omega_{x})\frac{\kappa a^{\xi-1}}{1+\kappa a^\xi}\rho_m$ (\citealt{Kaloper+etal+1998}). When $Q=0$ ($\xi+3\omega_{x}=0$) means DE does not interact with DM, and $Q\neq0$ ($\xi+3\omega_{x}\neq0$) means there is interaction. In addition, when $Q<0$ ($\xi+3\omega_{x}>0$) means the transfer of energy from DM to DE and the coincidence problem is not alleviated, when $Q>0$ ($\xi+3\omega_{x}<0$) indicates the conversion of DE to DM and the coincidence problem is alleviated.

Finally, we parameterize the Friedman equation to obtain:
\begin{equation}
E^{2}=a^{-3}\left[1-\Omega_{X, 0}\left(1-a^{\xi}\right)\right]^{-3 \omega_{x} / \xi}\label{2}.
\end{equation}
where $E=H/H_0$ is the dimensionless Hubble parameter.

The model has three parameters ($\Omega_{X,0},\omega_{x},\xi$) representing the current DE density parameter, equation of state of DE, and quantifying the severity of the coincidence problem. These parameters can be constrained by the following cosmological observations.

\begin{figure}[t]
\begin{center}
\includegraphics[width=0.6\textwidth,height=0.6\textwidth]{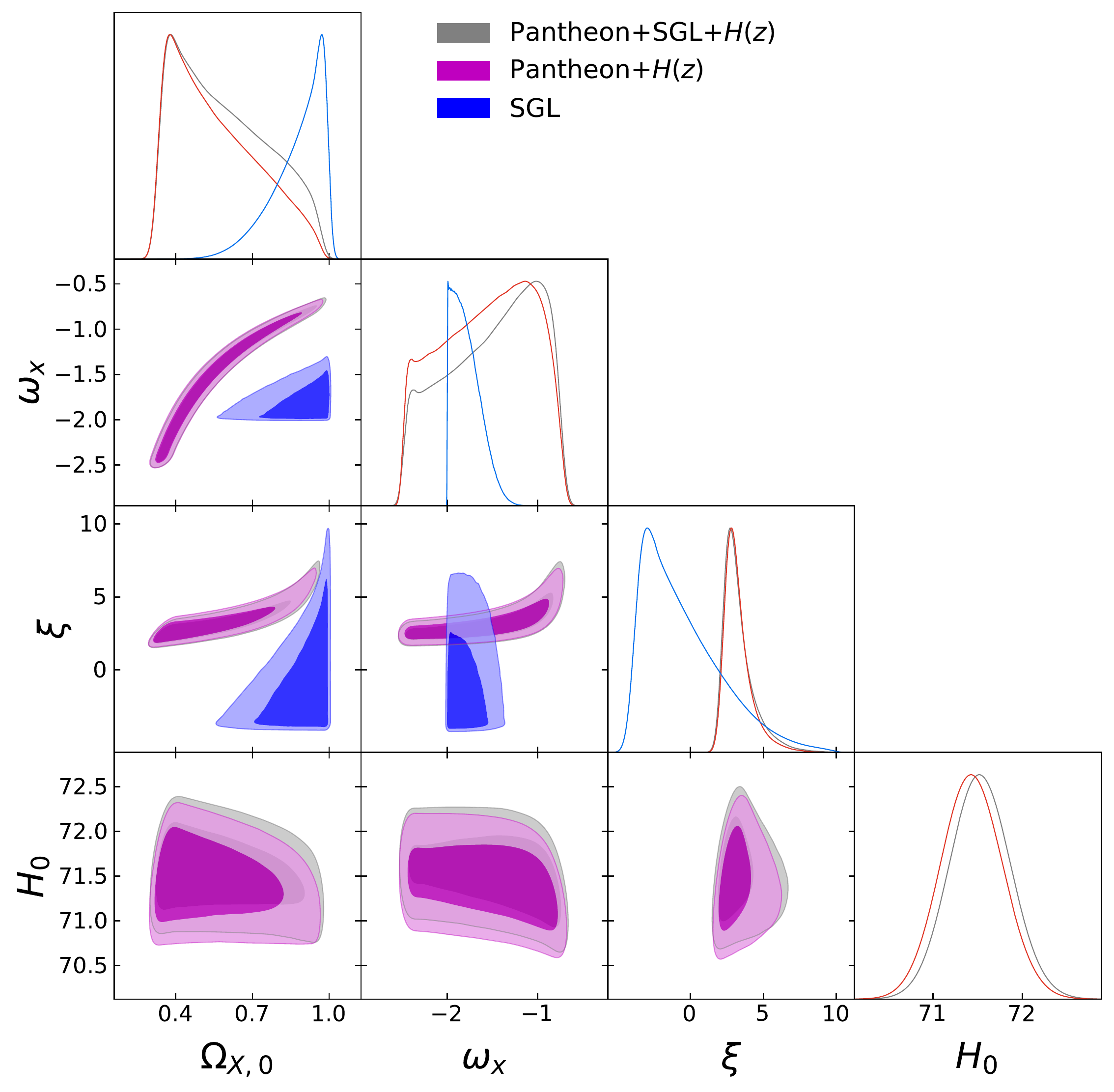}
\caption{Contour map of (Pantheon+SGL+$H(z)$), (Pantheon+$H(z)$) and (SGL) data combinations with constraints on model parameters ($\Omega_{X,0},\omega_x,\xi,H_0$). Note that since the SGL data has no constraint on $H_0$, the graph in the bottom row has only two contour plots of (Pantheon+SGL+$H(z)$) and (Pantheon+$H(z)$) constraint on $H_0$.}
\label{p2}
\end{center}
\end{figure}

\begin{table}
\begin{center}
\caption{Constraint results of (SGL), (Pantheon+$H(z)$) and (Pantheon+SGL+$H(z)$) data combinations on model parameters.}
\begin{tabular}{cccc}
\hline
Parameters                                 &~~SGL~~        &~~Pantheon+$H(z)$~~      &~~Pantheon+SGL+$H(z)$~~\\
\hline
$\Omega_{X,0}$      &$0.88\pm0.12$(1$\sigma$)$\pm0.21$(2$\sigma$)  &$0.56\pm0.22$(1$\sigma$)$\pm0.32$(2$\sigma$)&$0.59\pm0.25$(1$\sigma$)$\pm0.33$(2$\sigma$)\\
$\omega_x$        &$-1.78\pm0.21$(1$\sigma$)$\pm0.29$(2$\sigma$) &$-1.57\pm0.70$(1$\sigma$)$\pm0.88$(2$\sigma$)&$-1.50\pm0.72$(1$\sigma$)$\pm0.91$(2$\sigma$) \\
$\xi$               &$-0.5\pm3.3$(1$\sigma$)$\pm5.6$(2$\sigma$)     &$3.28\pm1.10$(1$\sigma$)$\pm2.10$(2$\sigma$)&$3.30\pm1.2$(1$\sigma$)$\pm2.4$(2$\sigma$)\\
$H_0$               &                                              &$71.44\pm0.35$(1$\sigma$)$\pm0.69$(2$\sigma$)&$71.53\pm0.34$(1$\sigma$)$\pm0.68$(2$\sigma$)\\
$\gamma_0$          &$1.23\pm0.08$(1$\sigma$)$\pm0.16$(2$\sigma$)  &  &$1.16\pm0.08$(1$\sigma$)$\pm0.15$(2$\sigma$) \\
$\gamma_z$          &$0.189\pm0.10$(1$\sigma$)$\pm0.19$(2$\sigma$)  &  &$-0.19\pm0.09$(1$\sigma$)$\pm0.17$(2$\sigma$) \\
$\gamma_s$         &$0.63\pm0.06$(1$\sigma$)$\pm0.11$(2$\sigma$)    &  &$0.66\pm0.059$(1$\sigma$)$\pm0.12$(2$\sigma$) \\
\hline
\end{tabular}
\label{22}
\end{center}
\end{table}

\section{Cosmological observation data and constraint results}
\label{sect:data}

In this section, we will introduce the three kinds of cosmological observational data used in this paper and the results of their constraints on parameters($\Omega_{X,0},\omega_{x},\xi$,$H_0$).

Firstly, we begin with the strong gravitational lensing (SGL) sample. In recent years, many new SGL systems have been discovered due to the emergence of powerful new telescopes for imaging and spectral observations. Therefore, it is of great practical significance to extract lens object information and cosmological parameters from SGL data. Hence, we use data from \citet{Chen+etal+2019} to constrain the parameters. They compiled 161 galaxy-scale sample systems of SGL. Due to the need for high-resolution HST imaging data, they eventually selected 130 galactic-scale SGL data from 161 samples. These 130 groups of data are from SLACS (57 data \citealt{Bolton+etal+2008,Auger+etal+2009,Auger+etal+2010}), S4TM (38 data \citealt{Shu+etal+2015,Shu+etal+2017}), BELLS (21 data \citealt{Brownstein+etal+2012}) and BELLS GALLERY (14 data \citealt{Shu+etal+2016,Shu+etal+2016a}), respectively.

We chose $R_{\text {eff }}/2$ as the radius because the half-light radius ($R_{\text {eff }}$) matches the Einstein radius well (\citealt{Auger+etal+2010}). From this, we can obtain observations of velocity dispersion:
\begin{equation}
\sigma_{\mathrm{obs}} =\sigma_{ap}\left[\theta_{\mathrm{eff}} /\left(2 \theta_{a p}\right)\right]^{\eta},
\end{equation}
where $\sigma_{ap}$ is velocity dispersion, $\theta_{\text {eff }}=R_{\text {eff }} / D_{l}$ and $D_{l}$ is the angular diameter distance, $\eta$ is the correction factor (\citealt{Cappellari+etal+2006}), and $\theta_{\mathrm{ap}} \approx 1.025 \times \sqrt{\left(\theta_{x} \theta_{y} / \pi\right)}$ with $\theta_{x}$ and $\theta_{y}$ being the angular sizes of width and length of the rectangular aperture, respectively (\citealt{Jorgensen+etal+1995}).

In addition, the theoretical expression of velocity dispersion is as follows:
\begin{equation}
\sigma_{th}=\sqrt{\frac{c^{2}}{2 \sqrt{\pi}} \frac{D_{s}}{D_{l s}} \theta_{E} \frac{3-\delta}{(\epsilon-2 \beta)(3-\epsilon)} F(\gamma, \delta, \beta)\left(\frac{\theta_{\mathrm{eff}}}{2 \theta_{\mathrm{E}}}\right)^{(2-\gamma)}}\label{3},
\end{equation}
where $D_{ls}$ is the angular diameter distance between lens and source, $D_s$ is the angular diameter distance of source, $\delta$ is the luminosity density slope and $\theta_{E}$ is the Einstein angle,  the $\epsilon= \gamma + \delta -2$ and $\beta$ is the orbit anisotropy parameter and it's annoying, we use Gaussian prior value $\beta=0.18$ for edge processing. Finally, the equation $F(\gamma, \delta, \beta)$ can be written in the form:
\begin{equation}\
F(\gamma, \delta, \beta)=\left[\frac{\Gamma[(\epsilon-1) / 2]}{\Gamma(\epsilon / 2)}-\beta \frac{\Gamma[(\epsilon+1) / 2]}{\Gamma[(\epsilon+2) / 2]}\right] \frac{\Gamma(\gamma / 2) \Gamma(\delta / 2)}{\Gamma[(\gamma-1) / 2] \Gamma[(\delta-1) / 2]},
\end{equation}
where $\gamma$ is total mass density slope and $\gamma=\gamma_0+\gamma_z z_l+\gamma_s \log \frac{\left(\sigma_{obs} /100~\mathrm{km}\mathrm{s}^{-1}\right)^{2}}{R_{\mathrm{eff}} / 10 h^{-1}~\mathrm{kpc}}$. Here, we consider the relationship between $\gamma$ and the redshift ($z_l$), surface mass density ($\sigma_{obs}^2/R_{eff}$). $\gamma_0$, $\gamma_z$ and $\gamma_s$ are the free parameter and we adopt the $H_0=100h~\rm{km}^{-1}\rm{Mpc}^{-1}$.

Secondly, it is well known that the observations of the type Ia supernovae (SNeIa) are direct evidence of the accelerated expansion of the universe (\citealt{Riess+etal+1998,Perlmutter+etal+1999}), and they have good constraints on the cosmological parameters. Since they are as bright as a typical galaxy at their peak, SNeIa can be seen at great distances, so that we can treat them as standard candle in cosmology. Additionally, this is the most effective and mature method for cosmological exploration. In this paper, we use the type Ia supernova data from \citet{Scolnic+etal+2018}. By using the effective distance estimation of various low-z and HST samples of SDSS SNLS, they formed 279 PS1 (Pan-STARRS1) SNeIa ($0.03<z<0.68$) into the largest SNeIa combination sample, with a total of 1048 SNeIa in the range $0.01<z<2.3$, which is called 'Pantheon sample'.

The distance modulus $\mu$ of a supernova is defined as follows:
\begin{equation}
\mu_{th} = m-M =5 \log (D_{l}/Mpc)+25,
\end{equation}
where $m$ is the  apparent magnitude, $M$ is the absolute magnitude and the luminosity distance $D_{l}$ is expressed as follows:
\begin{equation}
D_{l}=\frac{c (1+z)}{H_{0}} \int_{0}^{z} \frac{d z^{\prime}}{E\left(z^{\prime}\right)}.
\end{equation}
where $c$ is the speed of light.

For Pantheon sample, the observed distance modulus $\mu_{obs}^{Pantheon}$ (\citealt{Tripp+etal+1998}) can be given by:
\begin{equation}
\mu_{obs}^{Pantheon}=m_{B}-M+\varsigma y_{1}-\varrho n+\Delta M+\Delta B
\end{equation}
where $m_B$ and $M$ are the apparent magnitude and absolute magnitude of B-band, $y_{1}$ and $n$ are the light curve shape and colour parameter, $\varsigma$ and $\varrho$ are nuisance parameters. Furthermore, the $\Delta M$ and $\Delta B$ are the distance corrections and bias correction, respectively.

In addition, for the 580 type Ia supernova data (Union2.1) (\citealt{SupernovaCosmologyProject:2011ycw}) given by Supernova Cosmology Project (SCP), the observed distance modulus is:
\begin{equation}
\mu_{obs}^{Union2.1}=m_{B}-M+\varsigma_{1} x_{1}-\varrho_{1} c+\upsilon P\left(m_{true}<m_{threshold}\right)
\end{equation}
where $\varsigma_{1}$, $\varrho_{1}$ and $\upsilon$ are nuisance parameters, $P\left(m_{true}<m_{threshold}\right)$ is the integral of $P(m_{true})$ up to the threshold mass $m_{threshold}$. The $x_{1}$ and $c$ are the light curve shape and colour parameters.

Finally, we use the 31 Hubble parameters ($H(z)$) samples (\citealt{Wei+etal+2017}). The Hubble parameter has been widely used to constrain cosmology parameters in recent years. The Hubble parameter is a direct result of the zero-order kinetics of the universe, and it represents the expansion rate of the universe. In addition, among all cosmological measurements, the Hubble parameter is the only physical quantity that can directly measure the history of cosmic expansion. The advantage of the Hubble parameter is that it can be directly related to cosmic parameters without integration.

We use (SGL+SNe+$H(z)$) sample combination through the Monte Carlo Markov chain (MCMC) and minimum the $\chi^2$ to constrain the model parameters. The final $\chi_{All}^2$ is as follows:
\begin{equation}
\chi_{All}^{2}=\chi_{SGL}^{2}+\chi_{SNe}^{2}+\chi_{H(z)}^{2}.
\end{equation}
where $\chi_{SGL}^{2}$, $\chi_{SNe}^{2}$ and $\chi_{H(z)}^{2}$ can be expressed as follows:
\begin{equation}
\chi_{SGL}^{2}=\sum_{i=1}^{130}\left(\frac{\sigma_{\mathrm{th}}-\sigma_{\mathrm{obs}}}{\Delta \sigma_{\mathrm{tot}}}\right)^{2},
\end{equation}
\begin{equation}
\chi_{SNe}^{2}=\sum_{i=1}^{1048}\left(\frac{\mu_{th}-\mu_{obs}}{\sigma_{\mu}}\right)^{2},
\end{equation}
\begin{equation}
\chi_{H(z)}^{2}=\sum_{i=1}^{31}\left(\frac{H(z)_{th}-H(z)_{obs}}{\sigma_{H(z)}}\right)^{2}.
\end{equation}

\begin{figure}[t]
\begin{center}
\includegraphics[width=0.6\textwidth,height=0.6\textwidth]{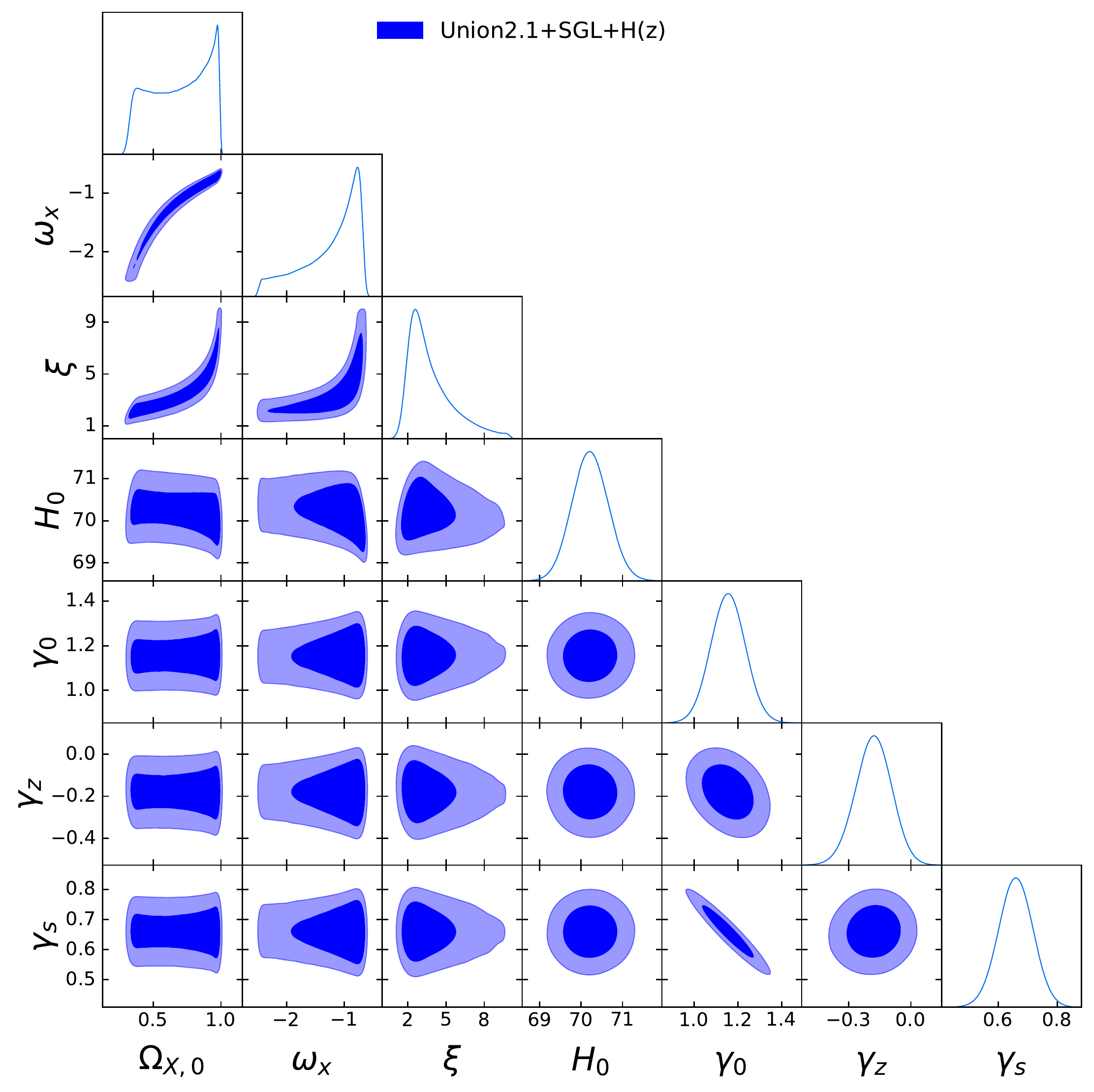}
\caption{Contour plot of the constraint of (Union2.1+SGL+$H(z)$) data combinations on model parameters.}
\label{p6}
\end{center}
\end{figure}

\begin{table}
\begin{center}
\caption{The constraint result of (Union2.1+SGL+$H(z)$) data on model parameters.}
\begin{tabular}{ccc}
\hline
Parameters                    &~~Union2.1+SGL+$H(z)$~~    \\
\hline
$\Omega_{X,0}$                &$0.69\pm0.20$(1$\sigma$)$\pm0.34$(2$\sigma$)\\
$\omega_x$                &$-1.24\pm0.59$(1$\sigma$)$\pm1.0$(2$\sigma$)\\
$\xi$                     &$3.8\pm2.1$(1$\sigma$)$\pm3.9$(2$\sigma$)\\
$H_0$                     &$70.22\pm0.43$(1$\sigma$)$\pm0.86$(2$\sigma$)\\
$\gamma_0$                &$1.16\pm0.08$(1$\sigma$)$\pm0.16$(2$\sigma$)\\
$\gamma_z$                &$-0.18\pm0.09$(1$\sigma$)$\pm0.17$(2$\sigma$) \\
$\gamma_s$                &$0.66\pm0.06$(1$\sigma$)$\pm0.11$(2$\sigma$)\\
\hline
\end{tabular}
\label{23}
\end{center}
\end{table}

In Table.\ref{22}, we show the results of the constraints on each model parameter by using three different data combinations, and the corresponding contour map is shown in Fig.\ref{p2}. The constraining power of SGL data on the parameter $\omega_x$ is stronger than that of (Pantheon+$H(z)$), and the error bar of $\omega_x=-1.78\pm0.21$ from SGL is about two times smaller than $\omega_x=-1.57\pm0.70$ from (Pantheon+$H(z)$). But for the other parameters, the constraint is not as strong as the (Pantheon+$H(z)$). By comparing the constraint results of (Pantheon+SGL+$H(z)$), we found that the addition of SGL data only had a weak impact on the results of (Pantheon+$H(z)$) and did not improve the constraint ability of data combination on parameters. We adopt the total constraint value $\xi=3.3\pm1.2$ and $\omega_x=-1.50\pm0.72$ and obtain $\xi+3\omega_x=-1.2\pm3.36$($1\sigma$), since the center value $\xi+3\omega_x=-1.2<0$ indicates that DE is converted to DM, the coincidence problem is slightly alleviated. However, it is clear that $\xi+3\omega_x=0$ is within the $1\sigma$ error range, which means that the data set (Pantheon+SGL+$H(z)$) does not fully distinguish between coincidence problem and non-coincidence problem. In addition, the measurement of $H_0$ in \citet{Riess+etal+2019} is $74.03\pm1.42$ kms$^{-1}$Mpc$^{-1}$ and the measurement of $H_0$ in \citet{Aghanim+etal+2020} is $67.4\pm0.5$ kms$^{-1}$Mpc$^{-1}$, and the deviation between these two results is $4.4\sigma$. However, the value of $H_0$ that we measured is $71.53\pm0.34$ kms$^{-1}$Mpc$^{-1}$ at the confidence interval of $1\sigma$, which deviate from the result of \citet{Aghanim+etal+2020} with $6.8\sigma$. Compared with the result of \citet{Riess+etal+2019} ($4.4\sigma$), we have more tension with the result of \citet{Aghanim+etal+2020}. This result shows that the Hubble tension problem does not disappear.

By considering the (Pantheon+$H(z)$) data may have a weak constraint on the parameter $\Omega_{X,0}$ in this model, which would affect the constraint on other parameters, we replace Pantheon data with Union2.1 data (\citealt{SupernovaCosmologyProject:2011ycw}) to constrain model parameters. The constraint results of the (Union2.1+SGL+$H(z)$) data combination are shown in Table.\ref{23} and Fig.\ref{p6}. At the confidence interval $1\sigma$, $\Omega_{X,0} = 0.69\pm0.20$, $\omega_x = -1.24\pm0.59$, $\xi = 3.8\pm2.1$ and $H_0 = 70.22\pm0.43$ kms$^{-1}$Mpc$^{-1}$. Compared with the parameter values given by the combination of the (Pantheon+SGL+$H(z)$) data, the accuracy of parameters $\xi$ and $H_0$ decreases by 75\% and 26\%, respectively. While the error accuracy of $\Omega_{X,0}$ and $\omega_x$ given by (Union2.1+SGL+$H(z)$) increases by 20\% and 18\% respectively. Therefore, it can be seen that the constraint precision of parameters $\xi$ and $H_0$ is decreased but the center of parameter $\Omega_{X,0}$ is worthy of optimization and the constraint precision is improved after Pantheon data is replaced by Union2.1 data. In addition, we adopt the total constraint value $\xi = 3.8\pm2.1$, $\omega_x = -1.24\pm0.59$ and obtain $\xi+3\omega_x=0.08\pm3.87$($1\sigma$), since the center value $\xi+3\omega_x=0.08>0$ indicates that DM is converted to DE, the problem of coincidences is not alleviated. However, we can see  that the $\xi+3\omega_x=0$ is within the $1\sigma$ error range, which means that the data set (Union2.1+SGL+$H(z)$) does not fully distinguish between coincidence problem and non-coincidence problem. Moreover, the value of $H_0$ is $70.22\pm0.43$ kms$^{-1}$Mpc$^{-1}$ at the confidence interval of $1\sigma$, which deviate from the result of \citet{Aghanim+etal+2020} with $4.3\sigma$. Compared with the results of (\citealt{Riess+etal+2019}) ($4.4\sigma$),  our results are slightly less in tension with \citet{Aghanim+etal+2020}. But the Hubble tension problem still does not disappear.
\begin{figure*}[htbp]
\begin{center}
\footnotesize
\begin{tabular}{ccc}
\includegraphics[width=0.5\textwidth]{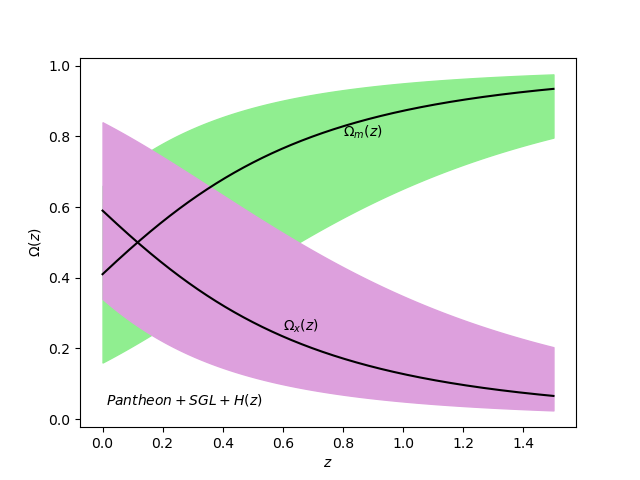} &
\includegraphics[width=0.5\textwidth]{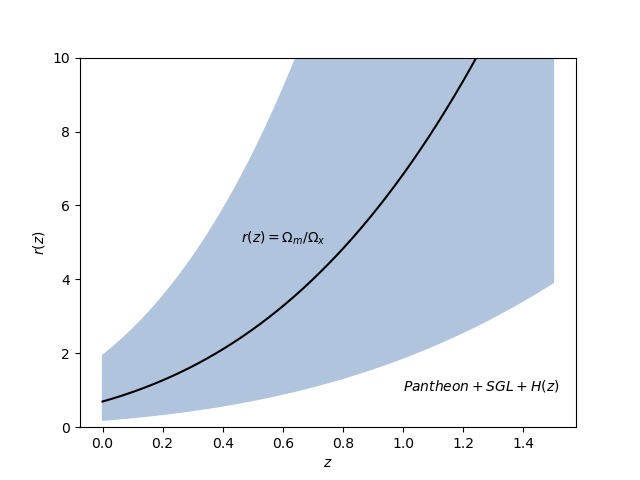} &\\
(a) & (b)\\
\includegraphics[width=0.5\textwidth]{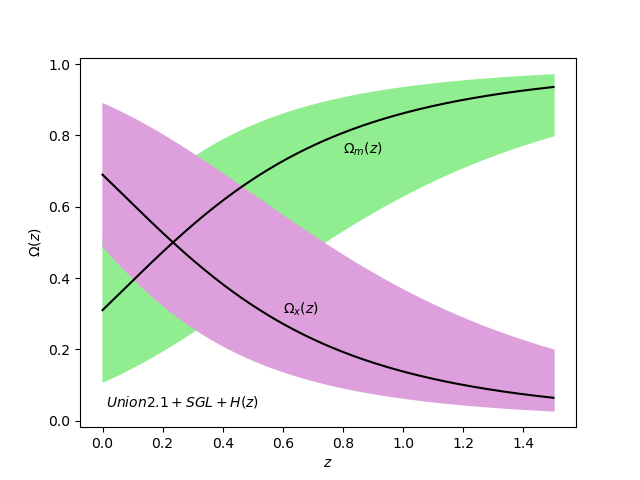} &
\includegraphics[width=0.5\textwidth]{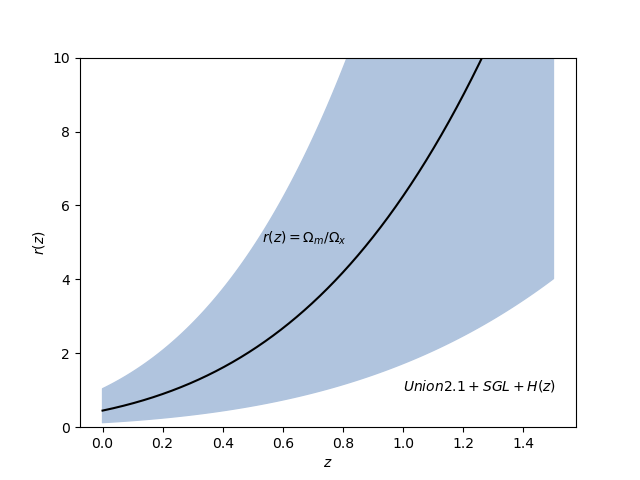} &\\
(c) & (d)\\
\end{tabular}
\end{center}
\caption{Figures (a) and (c) show the evolutions of $\Omega_m(z)$ and $\Omega_X(z)$ and the black line is the center value and the red line is the error value and figures (b) and (d) show the evolution of the ratio of the densities $r(z)=\Omega_m(z)/\Omega_X(z)$ respectively. }
\label{f2}
\end{figure*}

In addition, according to the formula $\Omega_{X}=\frac{\Omega_{X, 0} a^{\xi}}{1-\Omega_{X, 0}\left(1-a^{\xi}\right)}$, we show other constraint results of the combination of (Pantheon+SGL+$H(z)$) and (Union2.1+SGL+$H(z)$) data in Fig.\ref{f2}. Fig.\ref{f2} (a) and (c) shows the evolution of $\Omega_m$ and $\Omega_X$ with redshift, while Fig.\ref{f2} (b) and (d) shows the evolution of the ratio $r(z)$ of $\Omega_m$ and $\Omega_X$, indicating that they are have the same order of magnitude in $0<z<1.24$ (Pantheon+SGL+$H(z)$) and $0<z<1.26$ (Union2.1+SGL+$H(z)$). It should be noted that in order to simplify the calculation, we use the central value of $\xi$ and the error value of $\Omega_{X,0}$ in the range of $1\sigma$ to calculate the $1\sigma$ error of $\Omega(z)$, because we only consider the evolution trend of $\Omega(z)$ with $z$.

\section{The transition redshift $z_T$}
\label{sect:analysis}
\cite{Zhu+etal+2004} and \cite{Zhu+etal+2005} have proved that the transition redshift is an effective method to constrain the models. Based on Hubble parameters $H\equiv\dot{a}/a$ and deceleration parameters $q\equiv-\ddot{a}/\dot{a}^2$
\begin{equation}\label{4}
q=\left(-\frac{\ddot{a}/a}{H^{2}}\right)=\frac{d H^{-1}}{d t}-1,
\end{equation}
on account of $a=1/(1+z)$ and $E(z)=H/H_0$, we can rewrite Eq.~(\ref{4}) as
\begin{equation}\label{5}
q(z)=\frac{1}{2 E^{2}(z)} \frac{d E^{2}(z)}{d z}(1+z)-1,
\end{equation}
where $E^2(z)$ is Eq.~(\ref{2}). We can obtain the transition redshift $z_T$  by solving the equation.
\begin{equation}\label{6}
q=q(z=z_T)=0,
\end{equation}
From Eq.~(\ref{2}), Eq.~(\ref{5}), Eq.~(\ref{6}), we obtain the following solution
\begin{equation}\label{7}
\frac{1}{\left(1+z_{T}\right)^{\xi}}=\frac{\Omega_{X, 0}-1}{\Omega_{X, 0}\left(1+3 \omega_{X}\right)}.
\end{equation}

\begin{table}
\begin{center}
\caption{We compared the parameter values of (397SNe+BAO+CMB) data constraint in \citet{Chen+etal+2010} with the results obtained by the data used in this paper (Pantheon+SGL+$H(z)$) and (Union2.1+SGL+$H(z)$), it can be seen that our parameter value of $\Omega_{X,0}$ and $z_T$ are smaller than theirs.}
\begin{tabular}{cccc}
\hline
Data                    &~~$\Omega_{X,0}$~~   &~~$z_T$~~  &~~Reference~~\\
\hline
397SNe+BAO+CMB   &0.72  &0.73  &\citet{Chen+etal+2010}\\
Pantheon+SGL+$H(z)$     &0.58  &0.443 & present work\\
Union2.1+SGL+$H(z)$     &0.69  &0.645 & present work\\
\hline
\end{tabular}
\label{24}
\end{center}
\end{table}

\begin{figure*}[htbp]
\begin{center}
\footnotesize
\begin{tabular}{ccc}
\includegraphics[width=0.5\textwidth]{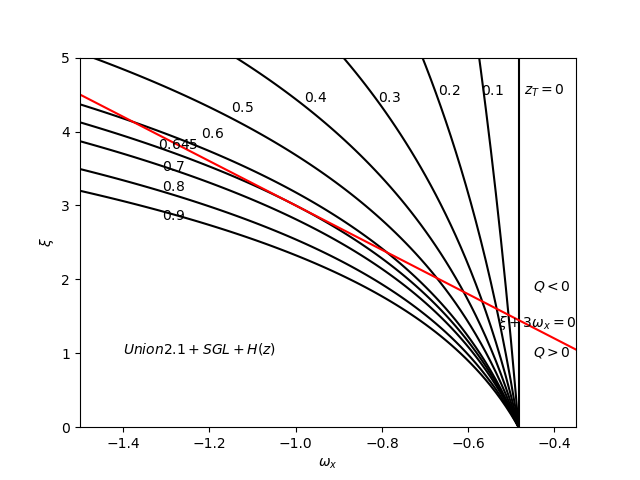} &
\includegraphics[width=0.5\textwidth]{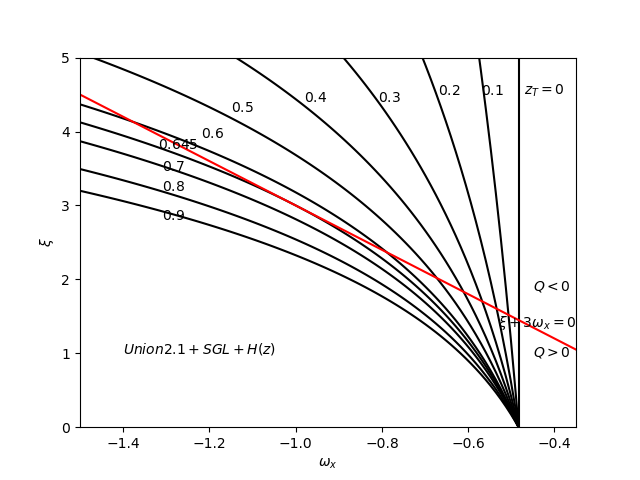} &\\
(a) & (b)\\
\end{tabular}
\end{center}
\caption{Figure (a) on the left is the function $z_T(\omega_x,\xi)$ diagram given by (Pantheon+SGL+$H(z)$) data and figure (b) on the right is given by (Union2.1+SGL+$H(z)$) data. The red line $\xi+3\omega_x =0$ represents no interaction between DE and DM.}
\label{f3}
\end{figure*}

We take the optimal value $\Omega_{X,0}=0.69$ of (Union2.1+SGL+$H(z)$) data constraint as a prior value. In Fig.\ref{f3} (b), we show the constraints of the transition redshift in $(\omega_x,\xi) $ plane. When the curve is tangent to the line $\xi+3\omega_x=0$, the coordinate of this point is $(\omega_x,\xi)=(-1.0,3.0)$ and $z_T=0.645$. Obviously, when $z_T>0.645$, the two lines do not intersect. Therefore, if the transition of the universe from a decelerated expansion state to an accelerated expansion state does occur at $z_T>0.645$, which means the interaction between DM and DE should be considered, and the energy is converted from DE to DM. On the other hand, if the transition occurs at $z_T<0.645$, just by using the transition redshift we can not guarantee the necessity of the interaction between DE and DM. In addition, Fig.\ref{f3} (a) shows the results of the transition redshift $z_T=0.443$ given by (Pantheon+SGL+$H(z)$) data. In Table.\ref{24}, we compared the parameter values of (397SNe+BAO+CMB) data constraint in \citet{Chen+etal+2010} with the results obtained by the data used in this paper (Pantheon+SGL+$H(z)$) and (Union2.1+SGL+$H(z)$), it can be seen that our parameter value of $\Omega_{X,0}$ and $z_T$ are smaller than theirs. In Amendola's work (\citealt{Amendola+etal+2003,Amendola+etal+2006}), they suggest that acceleration may start at a high redshift, or even reach $z\approx3$. If DE interacts strongly with DM, on the contrary, the standard noninteracting models are hard to achieve $z_T\simeq1$, which is consistent with our results.

\section{Discussions and conclusions}
\label{sect:discussion}
The interaction between DE and DM can be used to study the coincidence problem. In this work, we adopt the latest SGL data combined with $H(z)$ and SNeIa (Pantheon and Union2.1) data sets to constrain the phenomenological model. By using the MCMC method and minimum the $\chi^2$ we obtained the best constraint results. At the confidence interval of $1\sigma$, the conatraint results of (Pantheon+SGL+$H(z)$) data on the phenomenological model are $\Omega_{X,0}=0.59\pm0.25, \omega_x=-1.50\pm0.72$, $\xi=3.3\pm1.2$ and $\xi+3\omega_x=-1.20\pm3.36$ of which the center value indicates the coincidence problem is slightly alleviated but the $\xi+3\omega_x=0$ still within the $1\sigma$ error range. In addition, the value of $H_0 = 71.53\pm0.34$ kms$^{-1}$Mpc$^{-1}$ ($1\sigma$) can not relieve the tension problem of $H_0$. Furthermore, in order to test the constraint power of SGL data, we compare the results of three different data combinations (SGL, Pantheon+$H(z)$, Pantheon+SGL+$H(z)$). The results are shown in Table.\ref{22} and Fig.\ref{p2}. We found that the SGL data had a weak impact on the results of (Pantheon+$H(z)$) data.

Moreover, by considering the (Pantheon+$H(z)$) data may have a weak constraint on the parameter $\Omega_{X,0}$ in this model, we replace Pantheon data with Union2.1 data (\citealt{SupernovaCosmologyProject:2011ycw}) to constrain model parameters, and the results are shown in Table.\ref{23} and Fig.\ref{p6}. At the confidence interval of $1\sigma$, the constraint results of (Union2.1+SGL+$H(z)$) data on the phenomenological model are $\Omega_{X,0} = 0.69\pm0.20$, $\omega_x = -1.24\pm0.59$, $\xi = 3.8\pm2.1$ and $\xi+3\omega_x=0.08\pm3.87$ of which the center value indicates the coincidence problem is not alleviated but the $\xi+3\omega_x=0$ still within the $1\sigma$ error range. Moreover, the value of $H_0 = 70.22\pm0.43$ kms$^{-1}$Mpc$^{-1}$ ($1\sigma$) also can not relieve the tension problem of $H_0$. In addition, compared with the parameter values given by the combination of (Pantheon+SGL+$H(z)$) data, the constraint precision of parameters $\xi$ and $H_0$ is decreased, but the center of parameter $\Omega_{X,0}$ is worthy of optimization and the constraint precision is improved after Pantheon data is replaced by Union2.1 data.

Then we had two discussions, one is the evolution of DE and DM is consistent with the current observations, and the DE and DM have the same order of magnitude in the range of redshift $0<z<1.24$ (Pantheon+SGL+$H(z)$) and $0<z<1.26$ (Union2.1+SGL+$H(z)$), and the results were shown in Fig.\ref{f2}, the other is the theoretical constraint of the transition redshift suggests that if the universe changes from a decelerating expansion state to an accelerating expansion state, it occurs at a redshift $z_T>0.645$ and the result are shown in Fig.\ref{f3} (b), then in the phenomenological model, the interaction between DE and DM is necessary. In contrast, if this process occurs at a redshift $z_T<0.645$, it is impossible to determine the necessity of DE interacting with DM.

At last, the fact that the interaction is zero ($\xi+3\omega_x=0$) contained within $1\sigma$ indicates that the $\Lambda$CDM model is still the best fit for the observation, but it also indicates that current observations cannot clearly distinguish between standard cosmology and non-standard cosmology. We expect more strong gravitational lensing data, gravitational wave data and other observational data in the future to help us to research the cosmological issues.

\begin{acknowledgements}
This work is supported by the Chongqing Natural Science Foundation(Grants No.cstc2021jcyj-msxmX0553, Grants No.cstc2021jcyj-msxmX0481), Graduate Research and Innovation Foundation of Chongqing, China (Grant No. CYS21327).
\end{acknowledgements}

\bibliographystyle{raa}
\bibliography{RAA-2022-0125}

\label{lastpage}

\end{document}